# The transverse momentum distribution of $J/\psi$ mesons produced in $pp$ collisions at the LHC

Li-Na Gao*, Er-Qin Wang

Department of Physics, Taiyuan Normal University, Jinzhong, Shanxi 030619, China

**Abstract:** The transverse momentum distributions of $J/\psi$ mesons produced in $pp$ collisions at the center-of-mass energy 5 TeV, 7 TeV，and 13 TeV are described by the modified Hagedorn function. The fitting results by the modified Hagedorn function are in accord with experimental data measured by the LHCb Collaboration at LHC. The related parameters are obtained and analyzed.

1. Introduction

More and more scientific workers get involved in exploring the origin of the universe. Depending on modern cosmology, the quark-gluon plasma (QGP) is the initial state after the Big Bang and it is the original state of matter. So, exploring and studying the QGP is a way to understand the universe's evolution. A large number of research results show that the QGP may be existed in the extremely high temperature and high-density particle collision area. In the experiment, two antithetical heavy nuclei which speed approaches the light have a central collision to be formed a small "big bang". QGP may be created in the collision system. Creating and studying the QGP has become the main goal of the high energy heavy ion collision experiment. But, the new form of nuclear matter (QGP) cannot be directly observed by the detectors. We can only analyze the final particles of relativistic heavy ion collisions to infer the properties of QGP. To date, we have got a lot of experimental data which are provided by the Relativistic Heavy Ion Collider (RHIC) and the Large Hadron Collider (LHC) to study. And, scientists have achieved some achievements of the QCD diagram and the nuclear matter[1,2].

In the whole process of collision, the collision system is emitting particles and nuclear fragments constantly. The kinds and dynamical properties of the particles and nuclear fragments which emitted at different stages are not the same. Therefore, the final products of collisions have carried on a lot of the evolution information of collision[3]. The transverse momentum distribution of final state particles is an important object of observation in the experimentally, it can provide more important information to study the kinetic freezing temperature, the radial velocity of particles, chemical potential, the transverse excitation degree of collision system, and so on. At the high energy collision, the transverse momentum distribution of final state particles can be described well by the nonextensive Tsallis statistic distribution. In addition, theoretical researcher embedded different transverse expansion models into Tsallis statistic distributions to obtain more values of parameter[4-13], such as TBW model and Improved Tsallis distribution. TBW model is the blast wave model with Tsallis statistic distribution and Improved Tsallis distribution is the Tsallis statistic distribution with transverse flow effect. Using TBW model or Improved Tsallis distribution to describe the transverse momentum distribution of final state particles, we could get the average transverse (radial) flow velocity, nonextensivity parameter and the effective temperature. Transverse flow can also be embedded in Hagedorn function. The new function is called the modified Hagedorn function. In this paper, we used the modified Hagedorn function to describe the transverse momentum distribution of $J/\psi$ mesons produced in $pp$ collisions at different collision energy, extract and analyze some related parameters.

2. The model and method

In our previous work[14-18], we have revised some statistical thermodynamic distributions based on the multisource thermal model. In the work[14, 15], we put forward two components of statistical models which are the two-component Elang distribution and the two-component Schwinger mechanism. We used the two revised model to analyze the transverse momentum distributions of $\phi$ mesons, $\Omega$ hyperons, and negatively charged particles produced in Au-Au collisions with different centrality intervals, measured by the STAR Collaboration at 7.7 GeV, 11.5 GeV, 19.6 GeV, 27 GeV, and 39 GeV in the beam energy scan program at RHIC. And, we used the two methods to study the transverse momentum spectra of $J/\psi$ and $\Upsilon$ mesons produced in $pp$, $p$-Pb, Pb-Pb collisions measured by LHCb and ALICE Collaboration at LHC. In the work[16], we structured a two-component statistical model which is a superposition of the Tsallis statistics and the inverse power law. We used this two-component statistical model to analyze the transverse momentum distribution of $J/\psi$ and $\Upsilon$ mesons produced in $pp$, $p$-Pb collisions at 5 TeV, 7 TeV, 8 TeV, and 13 TeV measured by LHCb Collaboration at LHC. In the work[17], We used two models (the two-component Schwinger mechanism and the two-component statistical model which is based on the Tsallis statistics and the inverse power law) to study the $\Lambda_c^+$, $\Lambda_b^0$ baryons, $D^0$, $\bar{B}^0$ mesons and some related particles produced in $pp$, $p$-Pb collisions at 5 TeV, 7 TeV, 8 TeV, and 13 TeV. From our prophase related work, one can see that our revised models could describe the experimental data very well. In fitting the experimental data, we extracted some important related parameters and analyzed their trend with rapidity and collision energy. Through these studies work, we try to get some useful information about the collision mechanism, the evolution of the collision system and the QGP.

In this work, we used the modified Hagedorn function to study the transverse momentum distribution of $J/\psi$ mesons produced in $pp$ collisions at the center-of-mass energy 5 TeV, 7 TeV, and 13 TeV. To complete this paper, we introduced this function briefly in the following paragraphs.

In reference[19,20], the Hagedorn function is written as:

$$\frac{d^2N}{2\pi N_{ev} p_t dp_t dy} = C\left(1+\frac{m_t}{p_0}\right)^{-n} \quad (1)$$

In this function, $C$ is the fitting constant, $m_t = \sqrt{p_t^2 + m_0^2}$ denotes the transverse mass, $p_0$ and $n$ are the free parameters. Based on Quantum Chromodynamics, the Hagedorn function can be described the high transverse momentum part of transverse momentum distribution of hadrons very well.

It is well-known that the Tsallis function is a very useful method to study the transverse momentum of the final state particles in $pp$ collisions at RHIC and LHC energy range. The simplest form of the Tsallis function is[21,22]:

$$\frac{d^2N}{2\pi N_{ev} p_t dp_t dy} = C_q\left(1+(q-1)\frac{m_t}{T}\right)^{-1/(q-1)} \quad (2)$$

By comparing the Hagedorn function and Tsallis function, one can see that when the parameter $n$ can be expressed as $1/(q-1)$ and $p_0$ can be expressed as $nT$ ($T$ is the effective temperature). The Hagedorn function and the Tsallis function are mathematically equivalent. So, we could revise the function (1) to

$$\frac{d^2N}{2\pi N_{ev} p_t dp_t dy} = C\left(1+\frac{m_t}{nT_0}\right)^{-n} \quad (3)$$

Tsallis distribution has many different forms. In references[5,11,23,24], Tsallis distribution can be written as

$$E\frac{d^3N}{dp^3} = C_q\left(1+(q-1)\frac{E}{T}\right)^{-1/(q-1)} \quad (4)$$

Here $C_q$, $E$, $T$ and $q$ denote the normalization constant, particle energy, temperature and nonextensivity parameter, respectively. As stated earlier, we could let the parameter $n$ to replace $1/(q-1)$, the function (4) takes the form

$$E\frac{d^3N}{dp^3} = C_n\left(1+\frac{E}{nT}\right)^{-n}$$
$$= C_n\left(1+\frac{m_t}{nT}\right)^{-n} \quad (5)$$

As described in literature[23], the particle energy ($E$) can be replaced by the following four vector when particles' transverse flow exists in a co-moving system.

$$E = v^\mu p_\mu = \gamma\left(m_t - \vec{\beta}\cdot\vec{p}_T\right) \quad (6)$$

Where $v^\mu$ is four-velocity, $p_\mu$ is particles' four-momentum, and $v^\mu = \gamma(1,\vec{\beta},0)$, $p_\mu = (m_t,-\vec{p}_T,0)$, $\gamma = \frac{1}{\sqrt{1-\beta^2}}$. Hence, function (5) takes the form

$$E\frac{d^3N}{dp^3} = C_n\left(1+\frac{\gamma(m_t-\beta p_T)}{nT}\right)^{-n} \quad (7)$$

In this analysis, $m_t$ can be transform into $\langle\gamma_t\rangle(m_t - p_t\langle\beta_t\rangle)$, and $\langle\gamma_t\rangle = \frac{1}{\sqrt{1-\langle\beta_t\rangle^2}}$. In this way, function (3) can be written as

$$\frac{d^2N}{2\pi N_{ev} p_t dp_t dy} = C\left(1+\langle\gamma_t\rangle\frac{(m_t-p_t\langle\beta_t\rangle)}{nT_0}\right)^{-n} \quad (8)$$

Function (8) is the form of the modified Hagedorn function. We have used the function (8) to study the transverse momentum distribution of $J/\psi$ mesons produced in

$pp$ collisions at 5 TeV, 7 TeV, 8 TeV, and 13 TeV measured by LHCb Collaboration at LHC. The following section introduced the details of our research work.

## 3. Results and discussion

Figure 1 shows the transverse momentum distribution of $J/\psi$ mesons produced in $pp$ collisions at $\sqrt{s} =5$ TeV. Figure 1(a) presents the results of prompt $J/\psi$ mesons and Figure 1(b) present the results of nonprompt $J/\psi$ mesons, respectively. The hollow symbols with the error bars represent the experimental data measured by the LHCb Collaboration in literature[25], the different rapidity ranges are denoted by different symbols in the panels. In order to present clear representation of the transverse moment distribution, some distributions are scaled by factors. In the panels, different rapidity ranges are multiplied by different factors, such as in panel (a) 2.0<$y$<2.5 is multiplied by the factor $10^4$, 2.5<$y$<3.0 is multiplied by the factor $10^3$, 3.0<$y$<3.5 is multiplied by the factor $10^2$, 3.5<$y$<4.0 is multiplied by the factor $10^1$, 4.0<$y$<4.5 is multiplied by the factor $10^0$. The solid curves are our fitting results by the modified Hagedorn function. One can see that the modified Hagedorn function could describe the experimental data measured in $pp$ collisions at the center-of-mass energy 5 TeV by the LHCb Collaboration. Our fitting results are in agreement with the experimental data. During the fitting process, we extract the three fitting parameters ($n$, $T_0$, $\beta_t$). The values of the three fitting parameters and degree of freedom ($\chi^2/dof$) corresponding to each curve in Figure 1 are listed in Table 1. The trends of parameters on the rapidity will be discussed later.

Figure 2 and 3 are similar to Figure 1. Figure 2 show the transverse momentum spectra of 2(a) prompt $J/\psi$, 2(b) $J/\psi$ from $b$, 2(c) prompt $J/\psi$ with fully transversely polarized, 2(d) prompt $J/\psi$ with fully longitudinally polarized in $pp$ collisions at $\sqrt{s} =$ 7 TeV, respectively. Figures 3(a) and 3(b) show the results of prompt $J/\psi$ and $J/\psi$ from $b$ mesons in $pp$ collisions at $\sqrt{s} =13$TeV, respectively. The hollow symbols with the error bars represent the experimental data measured by the LHCb Collaboration in literature[26,27]. In the panels, different rapidity ranges are multiplied by different factors, and the solid curves are the results of our fitting by using the modified Hagedorn function. The values of the three fitting parameters ($n$, $T_0$, $\beta_t$) and degree of freedom ($\chi^2/dof$) corresponding to each curve in Figures 2 and 3 are also listed in Table 1, which will be discussed later. One can see that the fitting results by the modified Hagedorn function are fitted well with the experimental data.

In order to see clearly the relationship of the fitting parameters ($n$, $T_0$, $\beta_t$) and rapidity ($y$) for $J/\psi$ mesons produced in $pp$ collisions at 5 TeV, 7 TeV and 13 TeV, we plot the parameter values listed in Table 1 in Figure 4. In this Figure, the solid symbols are parameters and the solid lines are our fitting results by the least square method. The intercepts, slopes and $\chi^2/dof$ corresponding to each line are listed in Table 2. One can see that one parameter ($n$) is increased with the rapidity increase, and the other two parameters ($T_0$ and $\beta_t$) are decreased with the rapidity increase. It should be noted that the trend of $\beta_t$ has an abnormal phenomenon for $J/\psi$ mesons produced in $pp$ collisions at $\sqrt{s} = 7$ TeV.

In addition, we calculated mean $p_T$ ($\langle p_T \rangle$) and ratio of root-mean-square $p_T$ ($\sqrt{\langle p_T^2 \rangle}$) to $\sqrt{2}$ are also given in Table 1. Figure 5(a)-5(c) show the dependence of $\langle p_T \rangle$ on rapidity, 5(d)-5(f) show the dependence of $\sqrt{\langle p_T^2 \rangle/2}$ on rapidity, respectively. The solid symbols represent the values of $\langle p_T \rangle$ and $\sqrt{\langle p_T^2 \rangle/2}$ obtained from the curves in Figure 1-3. The solid lines are our fitting results by the least square method. The intercepts, slopes and $\chi^2/dof$ corresponding to each line are also listed in Table 2. One can see the decreasing trends of the $\langle p_T \rangle$ and $\sqrt{\langle p_T^2 \rangle/2}$ with the increase of rapidity. These decreasing trends of the $\langle p_T \rangle$ and $\sqrt{\langle p_T^2 \rangle/2}$ imply decreasing excitation degree of the interacting system with the increase of rapidity.

4. Summary and conclusions

Our main observations and conclusions are summarized as following:

(a) We described the transverse momentum distribution of $J/\psi$ mesons produced in $pp$ collisions at the center-of-mass energy 5 TeV, 7 TeV，and 13 TeV by the modified Hagedorn function. Our theoretical results agreed with the experimental data on the whole. These results suggest that the modified Hagedorn function is a good way to study the transverse momentum distribution of $J/\psi$ mesons produced in $pp$ collisions at LHC energy regions.

(b) We extracted the three free parameters ($n$, $T_0$, $\beta_t$) by fitting the experimental data. In order to obtain more information about the evolution of collision systems, we analyzed these parameters changes with rapidity at the different center-of-mass energy.

(c) Our research result show the parameter *n* is increased with the rapidity increase and not obviously different with the collision energy increased at LHC. As mentioned in literature[20], for point quark–quark scattering n ≈ 4, and the n parameter gets bigger when multiple scattering centers are involved[20,28-30]. Our result suggests that with the rapidity increased, the more multiple scattering centers are involved. And, the parameter *n* can be expressed as $1/(q-1)$ are described in section 2. The parameter *q* is the nonextensivity parameter, and represents the extent of nonthermalization. The increase of the parameter *n* means the decrease of the parameter *q*. Our result indicated that with the rapidity increased, the thermalization level of collision system larger.

(d) We calculated mean $p_T$ ($\langle p_T \rangle$) and ratio of root-mean-square $p_T$ to $\sqrt{2}$ ($\sqrt{\langle p_T^2 \rangle/2}$). And, we analyzed the behaviors of $\langle p_T \rangle$ and $\sqrt{\langle p_T^2 \rangle/2}$. The two considered quantities ($\langle p_T \rangle$ and $\sqrt{\langle p_T^2 \rangle/2}$) decrease with the increase of rapidity. These decreasing trends of the $\langle p_T \rangle$ and $\sqrt{\langle p_T^2 \rangle/2}$ imply decreasing excitation degree of the interacting system with the increase of rapidity. And, the initial temperature of the interacting system is approximately described by $\sqrt{\langle p_T^2 \rangle/2}$ are expressed in reference[31-33]. So, the initial temperature of the interacting system decreases when the rapidity increases. It is mainly because of less energy deposition in very forward rapidity region.

(e) As mentioned in section 2, $\beta_t$ is the average transverse (radial) flow velocity. And, $T_0$ is an estimate for the kinetic freeze-out temperature. The two parameters ($T_0$ and $\beta_t$) are decreased with the rapidity increase. This result means that the collision system may be having faster expansion and higher excitation degree in low rapidity region. In addition, it must be noticed that the trend of $\beta_t$ has an abnormal phenomenon for *J/ψ* mesons produced in *pp* collisions at $\sqrt{s} = 7$TeV. This special phenomenon deserves our attention. Our next step work is exploring the reasons.

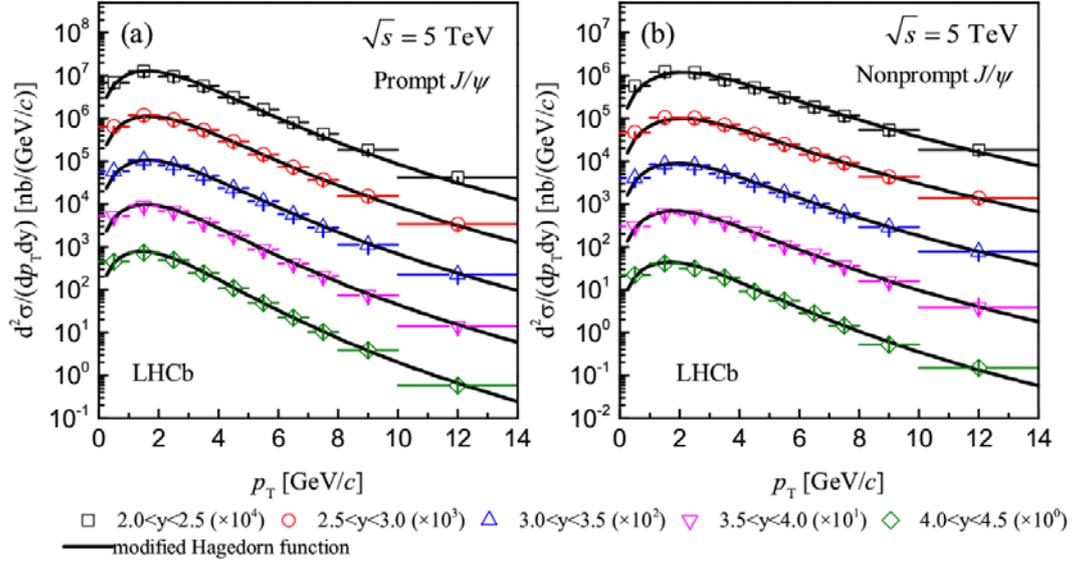

Figure 1: Transverse momentum distribution of (a) prompt $J/\psi$ (b) noprompt $J/\psi$ mesons produced in $pp$ collisions at $\sqrt{s}=5$ TeV. The hollow symbols with the error bars represent the experimental data of the LHCb Collaboration in literature[25], the curves are our results calculated by the modified Hagedorn function.

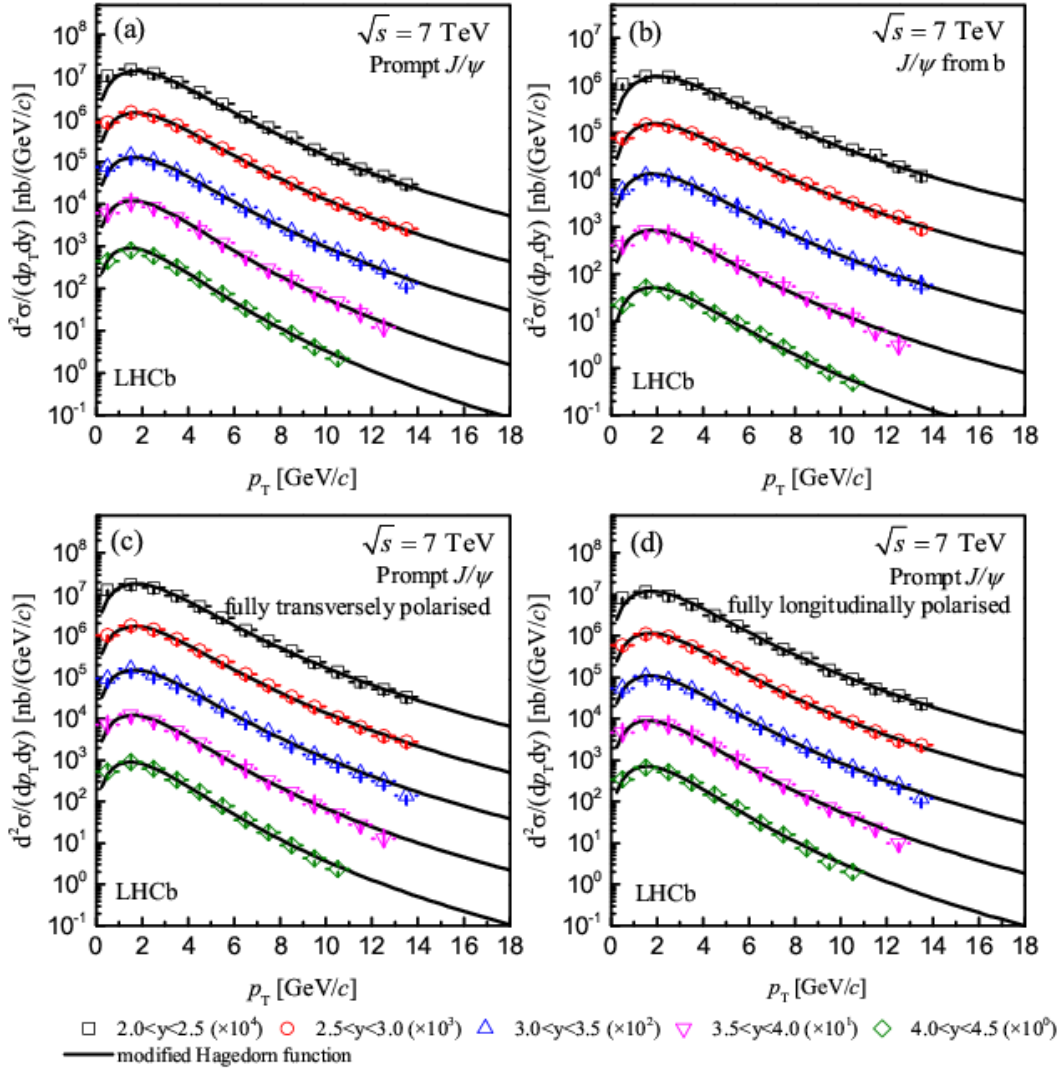

Figure 2: The same as Figure 1, but showing the results of(a) prompt $J/\psi$ (b) $J/\psi$ from b (c) prompt $J/\psi$ (assuming fully transversely polarised) (d)prompt $J/\psi$ (assuming fully longitudinally polarised) mesons produced in *pp* collisions at $\sqrt{s}$ =7 TeV. The experimental data are quoted from the literature[26].

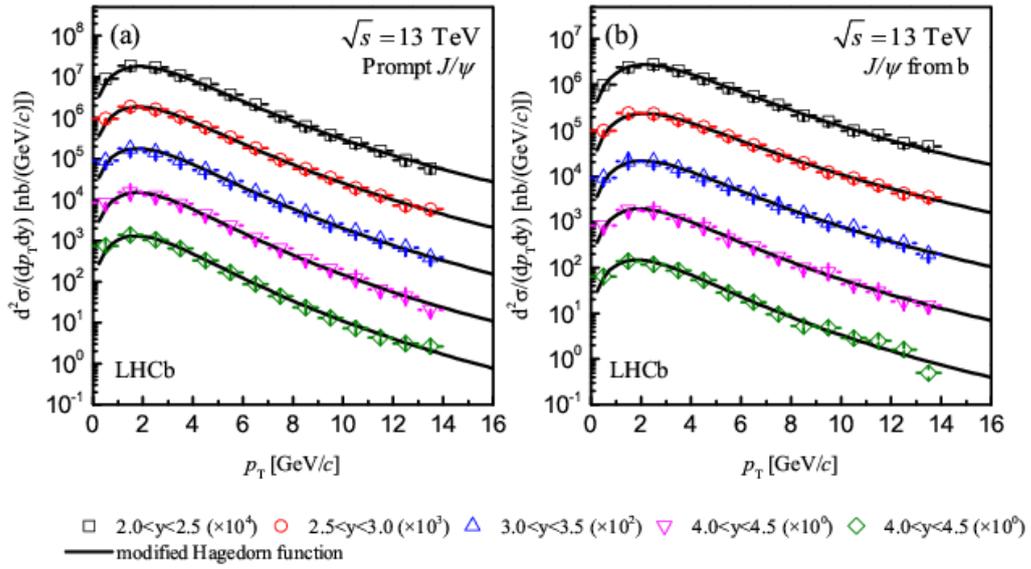

Figure 3: The same as Figure 1, but showing the results of (a) prompt $J/\psi$ (b) $J/\psi$ from b mesons produced in $pp$ collisions at $\sqrt{s}$ = 13 TeV. The experimental data are quoted from the literature[27].

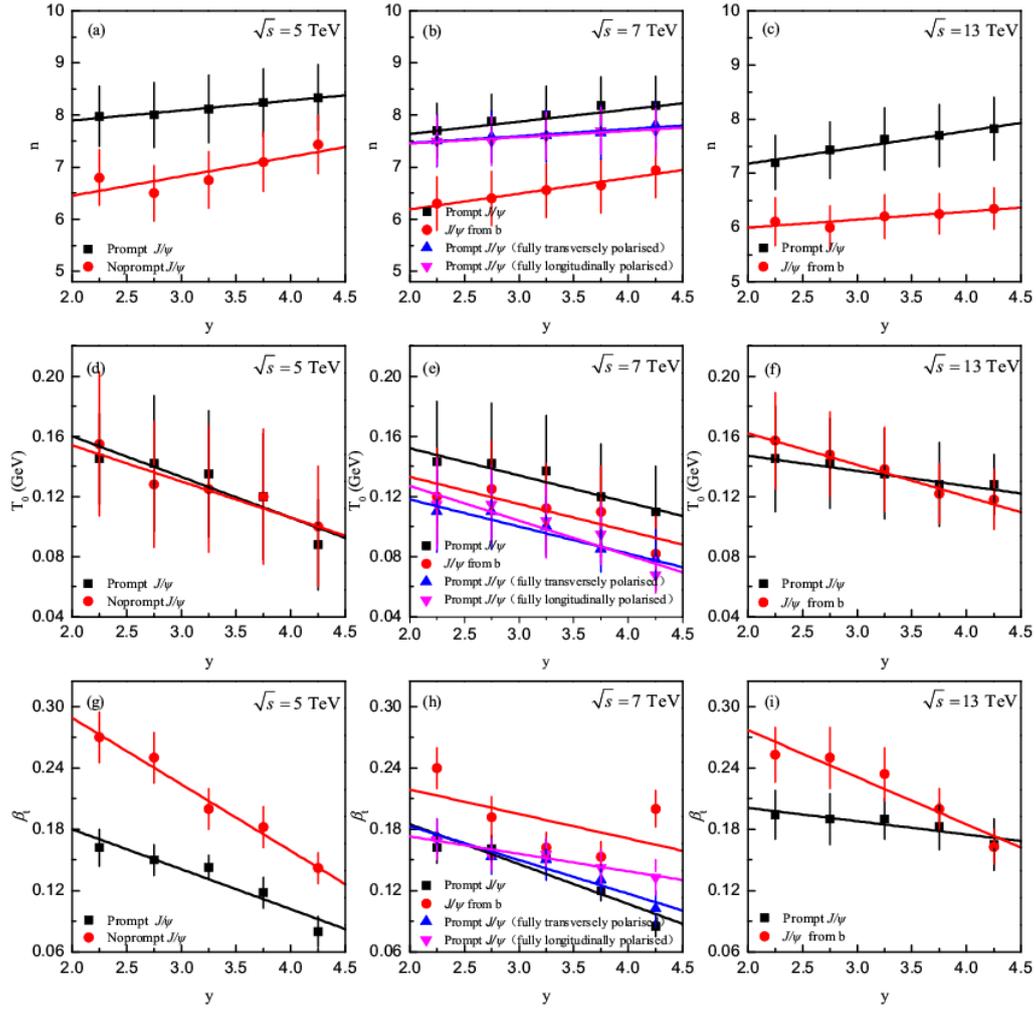

Figure 4: The relationship of free parameters ($n$, $T_0$, $\beta_t$) and rapidity($y$) for $J/\psi$ mesons produced in $pp$ collisions at 5 TeV, 7 TeV, and 13TeV. The solid symbols are quoted in Table 2, and the lines are our fitted results by the least square method.

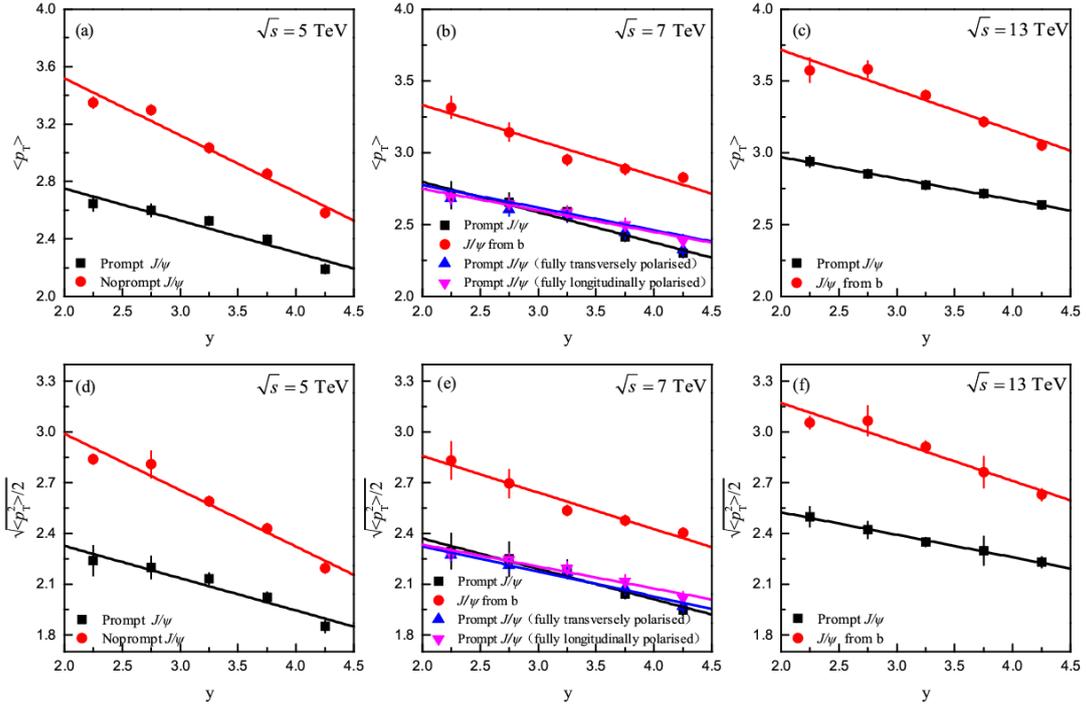

Figure 5: The same as Figure 4, but showing the relationship between physical quantity ((a)-(c) $\langle p_T \rangle$, (d)-(f) $\sqrt{\frac{\langle p_T^2 \rangle}{2}}$ )and $y$ for $J/\psi$ mesons produced in $pp$ collisions at 5 TeV, 7 TeV, and 13 TeV.

Table 1: Values of parameters and $\chi^2/dof$ corresponding to the curves in Figure 1-3.

| Figure | Type | $n$ | $T_0$(GeV) | $\beta_t$ | $\chi^2/dof$ | $\langle p_T \rangle$(GeV/c) | $\sqrt{\langle p_T^2 \rangle/2}$(GeV/c) |
|---|---|---|---|---|---|---|---|
| Figure 1(a) | 2.0<y<2.5 | 7.975±0.580 | 0.145±0.030 | 0.162±0.018 | 1.875 | 2.647±0.051 | 2.238±0.088 |
|  | 2.5<y<3.0 | 8.000±0.620 | 0.142±0.045 | 0.150±0.015 | 0.751 | 2.600±0.042 | 2.199±0.067 |
|  | 3.0<y<3.5 | 8.120±0.650 | 0.135±0.042 | 0.143±0.012 | 0.694 | 2.526±0.023 | 2.134±0.033 |
|  | 3.5<y<4.0 | 8.243±0.650 | 0.120±0.042 | 0.118±0.015 | 0.402 | 2.397±0.014 | 2.024±0.028 |
|  | 4.0<y<4.5 | 8.335±0.635 | 0.088±0.030 | 0.080±0.015 | 0.357 | 2.191±0.032 | 1.849±0.034 |
| Figure 1(b) | 2.0<y<2.5 | 6.800±0.530 | 0.155±0.048 | 0.270±0.025 | 1.467 | 3.350±0.038 | 2.840±0.027 |
|  | 2.5<y<3.0 | 6.500±0.530 | 0.128±0.042 | 0.250±0.025 | 1.672 | 3.298±0.033 | 2.809±0.078 |
|  | 3.0<y<3.5 | 6.752±0.543 | 0.125±0.042 | 0.200±0.020 | 0.379 | 3.036±0.027 | 2.592±0.026 |
|  | 3.5<y<4.0 | 7.102±0.570 | 0.120±0.045 | 0.182±0.020 | 0.333 | 2.852±0.029 | 2.428±0.027 |
|  | 4.0<y<4.5 | 7.433±0.565 | 0.100±0.040 | 0.142±0.015 | 0.883 | 2.584±0.032 | 2.196±0.031 |
| Figure 2(a) | 2.0<y<2.5 | 7.703±0.522 | 0.143±0.040 | 0.162±0.015 | 1.256 | 2.706±0.093 | 2.294±0.104 |
|  | 2.5<y<3.0 | 7.880±0.520 | 0.142±0.040 | 0.161±0.013 | 0.352 | 2.657±0.064 | 2.248±0.100 |
|  | 3.0<y<3.5 | 8.000±0.550 | 0.137±0.037 | 0.152±0.013 | 0.460 | 2.590±0.033 | 2.189±0.053 |
|  | 3.5<y<4.0 | 8.182±0.550 | 0.120±0.035 | 0.120±0.010 | 0.312 | 2.414±0.013 | 2.040±0.012 |
|  | 4.0<y<4.5 | 8.184±0.556 | 0.110±0.030 | 0.085±0.010 | 0.173 | 2.300±0.024 | 1.947±0.013 |
| Figure 2(b) | 2.0<y<2.5 | 6.303±0.517 | 0.120±0.032 | 0.240±0.020 | 1.490 | 3.316±0.075 | 2.833±0.109 |
|  | 2.5<y<3.0 | 6.400±0.520 | 0.125±0.032 | 0.192±0.020 | 0.277 | 3.143±0.062 | 2.696±0.082 |
|  | 3.0<y<3.5 | 6.555±0.520 | 0.112±0.030 | 0.162±0.015 | 0.153 | 2.951±0.037 | 2.534±0.021 |
|  | 3.5<y<4.0 | 6.648±0.520 | 0.110±0.030 | 0.153±0.015 | 0.483 | 2.886±0.035 | 2.477±0.019 |
|  | 4.0<y<4.5 | 6.935±0.525 | 0.082±0.025 | 0.200±0.018 | 0.147 | 2.827±0.021 | 2.403±0.014 |
| Figure 2(c) | 2.0<y<2.5 | 7.505±0.500 | 0.110±0.027 | 0.173±0.017 | 0.823 | 2.679±0.035 | 2.270±0.013 |
|  | 2.5<y<3.0 | 7.583±0.500 | 0.110±0.025 | 0.153±0.017 | 0.440 | 2.604±0.028 | 2.208±0.028 |
|  | 3.0<y<3.5 | 7.602±0.503 | 0.100±0.020 | 0.150±0.020 | 0.690 | 2.559±0.020 | 2.169±0.012 |
|  | 3.5<y<4.0 | 7.660±0.510 | 0.085±0.015 | 0.130±0.015 | 0.578 | 2.445±0.022 | 2.072±0.012 |
|  | 4.0<y<4.5 | 7.801±0.450 | 0.078±0.020 | 0.102±0.018 | 0.227 | 2.322±0.018 | 1.967±0.010 |
| Figure 2(d) | 2.0<y<2.5 | 7.503±0.477 | 0.115±0.030 | 0.170±0.020 | 0.706 | 2.688±0.025 | 2.279±0.036 |
|  | 2.5<y<3.0 | 7.505±0.470 | 0.115±0.025 | 0.155±0.015 | 0.158 | 2.646±0.023 | 2.246±0.029 |
|  | 3.0<y<3.5 | 7.600±0.450 | 0.104±0.025 | 0.156±0.015 | 0.254 | 2.588±0.018 | 2.193±0.011 |
|  | 3.5<y<4.0 | 7.702±0.450 | 0.095±0.020 | 0.143±0.017 | 0.945 | 2.500±0.015 | 2.116±0.013 |
|  | 4.0<y<4.5 | 7.706±0.474 | 0.068±0.012 | 0.133±0.017 | 0.283 | 2.386±0.019 | 2.018±0.014 |
| Figure 3(a) | 2.0<y<2.5 | 7.202±0.500 | 0.145±0.035 | 0.194±0.024 | 0.950 | 2.940±0.037 | 2.500±0.059 |
|  | 2.5<y<3.0 | 7.430±0.520 | 0.142±0.030 | 0.190±0.025 | 0.849 | 2.853±0.022 | 2.421±0.049 |
|  | 3.0<y<3.5 | 7.632±0.578 | 0.135±0.030 | 0.190±0.020 | 0.799 | 2.777±0.012 | 2.349±0.018 |
|  | 3.5<y<4.0 | 7.701±0.580 | 0.128±0.028 | 0.183±0.023 | 1.703 | 2.716±0.005 | 2.296±0.084 |
|  | 4.0<y<4.5 | 7.820±0.580 | 0.128±0.020 | 0.165±0.025 | 1.578 | 2.637±0.003 | 2.230±0.031 |
| Figure 3(b) | 2.0<y<2.5 | 6.107±0.443 | 0.157±0.032 | 0.253±0.027 | 0.439 | 3.574±0.084 | 3.055±0.037 |
|  | 2.5<y<3.0 | 6.000±0.400 | 0.148±0.028 | 0.250±0.030 | 0.540 | 3.581±0.055 | 3.065±0.088 |
|  | 3.0<y<3.5 | 6.203±0.403 | 0.138±0.028 | 0.234±0.026 | 0.608 | 3.402±0.021 | 2.912±0.033 |
|  | 3.5<y<4.0 | 6.252±0.372 | 0.122±0.020 | 0.200±0.020 | 0.529 | 3.217±0.026 | 2.762±0.091 |
|  | 4.0<y<4.5 | 6.350±0.380 | 0.118±0.020 | 0.163±0.017 | 1.909 | 3.053±0.022 | 2.629±0.036 |

Table 2: Values of intercepts, slopes, and $\chi^2/dof$ corresponding to the lines in Figures 4 and 5.

| Figure | Type | Intercept | Slope | $\chi^2/dof$ |
|---|---|---|---|---|
| Figure 4(a) | prompt $J/\psi$ | 7.509±0.057 | 0.193±0.017 | 0.003 |
| | noprompt$J/\psi$ | 5.703±0.429 | 0.374±0.129 | 0.193 |
| Figure 4(b) | prompt $J/\psi$ | 7.168±0.101 | 0.253±0.030 | 0.010 |
| | $J/\psi$ from b | 5.585±0.111 | 0.302±0.033 | 0.014 |
| | prompt $J/\psi$(fully transversely polarized) | 7.195±0.067 | 0.134±0.020 | 0.006 |
| | prompt $J/\psi$(fully longitudinally polarized) | 7.211±0.064 | 0.121±0.019 | 0.006 |
| Figure 4(c) | prompt $J/\psi$ | 6.577±0.108 | 0.301±0.033 | 0.012 |
| | $J/\psi$ from b | 5.703±0.141 | 0.148±0.042 | 0.035 |
| Figure 4(d) | prompt $J/\psi$ | 0.214±0.019 | -0.027±0.006 | 0.095 |
| | noprompt$J/\psi$ | 0.202±0.014 | -0.024±0.004 | 0.029 |
| Figure 4(e) | prompt $J/\psi$ | 0.188±0.010 | -0.018±0.003 | 0.021 |
| | $J/\psi$ from b | 0.169±0.018 | -0.018±0.005 | 0.116 |
| | prompt $J/\psi$(fully transversely polarized) | 0.154±0.008 | -0.018±0.002 | 0.041 |
| | prompt $J/\psi$(fully longitudinally polarized) | 0.173±0.015 | -0.023±0.005 | 0.252 |
| Figure 4(f) | prompt $J/\psi$ | 0.167±0.004 | -0.010±0.001 | 0.008 |
| | $J/\psi$ from b | 0.204±0.005 | -0.021±0.002 | 0.020 |
| Figure 4(g) | prompt $J/\psi$ | 0.258±0.021 | -0.039±0.006 | 0.684 |
| | noprompt$J/\psi$ | 0.419±0.015 | -0.065±0.005 | 0.145 |
| Figure 4(h) | prompt $J/\psi$ | 0.263±0.026 | -0.039±0.008 | 1.336 |
| | $J/\psi$ from b | 0.267±0.061 | -0.024±0.018 | 3.890 |
| | prompt $J/\psi$(fully transversely polarized) | 0.249±0.014 | -0.033±0.004 | 0.172 |
| | prompt $J/\psi$(fully longitudinally polarized) | 0.207±0.007 | -0.017±0.002 | 0.072 |
| Figure 4(i) | prompt $J/\psi$ | 0.227±0.011 | -0.013±0.003 | 0.071 |
| | $J/\psi$ from b | 0.369±0.025 | -0.046±0.007 | 0.339 |
| Figure 5(a) | prompt $J/\psi$ | 3.197±0.106 | -0.223±0.032 | 5.515 |
| | noprompt$J/\psi$ | 4.310±0.122 | -0.396±0.037 | 3.867 |
| Figure 5(b) | prompt $J/\psi$ | 3.219±0.079 | -0.211±0.024 | 1.735 |
| | $J/\psi$ from b | 3.827±0.106 | -0.247±0.032 | 3.345 |
| | prompt $J/\psi$(fully transversely polarized) | 3.089±0.052 | -0.157±0.016 | 1.976 |
| | prompt $J/\psi$(fully longitudinally polarized) | 3.049±0.049 | -0.150±0.015 | 1.941 |
| Figure 5(c) | prompt $J/\psi$ | 3.268±0.014 | -0.149±0.004 | 0.639 |
| | $J/\psi$ from b | 4.279±0.121 | -0.281±0.036 | 2.582 |
| Figure 5(d) | prompt $J/\psi$ | 2.708±0.090 | -0.191±0.027 | 1.809 |
| | noprompt$J/\psi$ | 3.658±0.115 | -0.334±0.035 | 3.347 |
| Figure 5(e) | prompt $J/\psi$ | 2.730±0.063 | -0.180±0.019 | 1.212 |
| | $J/\psi$ from b | 3.290±0.072 | -0.216±0.022 | 3.840 |
| | prompt $J/\psi$(fully transversely polarized) | 2.619±0.045 | -0.148±0.013 | 4.597 |
| | prompt $J/\psi$(fully longitudinally polarized) | 2.594±0.043 | -0.130±0.013 | 2.625 |
| Figure 5(f) | prompt $J/\psi$ | 2.791±0.015 | -0.133±0.005 | 0.121 |
| | $J/\psi$ from b | 3.635±0.101 | -0.231±0.030 | 1.463 |

**Data Availability**

All data are quoted from the mentioned references. As a phenomenological work, this paper does not report new data.

**Conflicts of Interest**

The authors declare that there are no conflicts of interest regarding the publication of this paper.

**Acknowledgments**

Author Li-Na Gao acknowledges the financial support from the National Natural Science Foundation of China under Grant No. 11847003, the Shanxi Provincial Science and Technology Innovation Plan under Grant No. 2019L0804, theuniversity student innovation and entrepreneurship trainingprogram of Taiyuan Normal University No. CXCY 2276, the Doctoral Scientific Research Foundation of Taiyuan Normal University under Grant No. I170167, and the Doctoral Scientific Research Foundation of Shanxi Province under Grant No. I170269.